# Spin transport in self assembled all-metal nanowire spin valves: A study of the pure Elliott-Yafet mechanism

Sandipan Pramanik[*], Carmen-Gabriela Stefanita and Supriyo Bandyopadhyay
Department of Electrical and Computer Engineering
Virginia Commonwealth University
Richmond, VA 23284, USA

**Abstract:** We report experimental study of spin transport in all metal nanowire spin valve structures. The nanowires have a diameter of 50 nm and consist of three layers – cobalt, copper and nickel. Based on the experimental observations, we determine that the primary spin relaxation mechanism in the paramagnet copper is the Elliott-Yafet mode associated with frequent interface roughness scattering. This mode is overwhelmingly dominant over all other modes, so that we are able to study the pure Elliott-Yafet mechanism. We deduce that the spin diffusion length associated with this mechanism is about 16 nm in our nanowires and is fairly temperature independent in the range 1- 100 K. The corresponding spin relaxation time is about 100 femtoseconds. We also find that the spin relaxation is fairly independent of the electric field driving the current in the field range 0.75 – 7.5 kV/cm.

## 1. INTRODUCTION

All-metal spin valve devices with Cu as the spacer material have been studied in various configurations[1, 2] by several groups in the past. However, the dominant spin relaxation mechanism in the Cu spacer was never conclusively identified. There exist four possible mechanisms that can cause spin relaxation in Cu. These are (1) D'yakonov-Perel' (DP) mechanism[3], (2) Elliott-Yafet (EY) mechanism[4], (3) Bir-Aronov-Pikus (BAP) mechanism[5] and (4) hyperfine interaction (HFI) between nuclear and carrier (electron/hole) spins[6]. The most likely candidates for spin relaxation in Cu are the DP and EY processes because the BAP mechanism is absent for unipolar transport and the HFI mechanism is much weaker compared to the DP and the EY in most solids, including Cu.

Neither the DP nor the EY relaxation rate has been individually measured in Cu since it is not normally possible to separate the two. However, there is a way to make one rate much higher than the other. We measure the spin relaxation length in a *nanowire* spin valve consisting of Co-Cu-Ni as shown in Fig. 1. In a nanowire, electrons will experience increased surface roughness scattering which will decrease carrier mobility. The DP mechanism is suppressed by frequent elastic scattering[3], but the Elliott-Yafet mechanism is enhanced[4]. Thus, confining carriers in a nanowire will make the EY rate overwhelmingly dominant over the DP rate and thus allow us to measure the relaxation rate due to the EY mechanism alone. This, then, will allow us to probe various features of the EY mechanism in a metal, such as the temperature and electric field dependence of the EY relaxation rate.

## 2. EXPERIMENT

In order to fabricate a nanowire spin valve structure, we start with a high purity (99.997%) metallic aluminum foil (0.1 mm thick), which is electropolished in a suitable organic solution[7] to produce a mirror like surface. An anodic alumina film with highly ordered nanopores is formed on this electropolished surface by a multistep anodization procedure[8]. The anodization conditions (e.g. the nature of the acidic electrolyte, anodization voltage, duration of final step anodization etc.) determine the dimensions of the nanopores. In this work we have used 0.3M oxalic acid as the electrolyte and anodization voltage has been kept constant at 40V dc. Under these conditions we get a porous alumina film with nominal pore diameter of 50 nm (Fig. 2 (a)). Final step anodization was carried out for 90 seconds which makes the pores ~ 100 nm deep. The insulating alumina barrier layer at the pore bottom is removed by a "reverse polarity etching" technique[9], which results in a "through hole" nanopattern on the bulk aluminum (Fig. 2 (b)). This facilitates dc electrodeposition of metals inside the pores while retaining the bulk aluminum substrate at the back of the template.

A spin-valve configuration is a trilayered structure in which a nonmagnetic layer is sandwiched between two ferromagnetic electrodes. In this work we choose the nonmagnetic "spacer" layer to be made of Cu and the ferromagnetic electrodes are chosen to be Co and Ni. These trilayered quantum wires are self assembled in the nanoporous alumina matrix by sequentially electrodepositing Ni, Cu and Co inside the pores. Electrodeposition has been carried out by applying a small dc bias of 1.5V at a platinum counter electrode with respect to the aluminum foil. The electrolyte is a dilute aqueous solution of the metal-sulfate salt with slightly acidic pH. Small deposition current (~ $\mu$A) ensures slow and well-controlled electrodeposition of metals inside the pores. We calibrated the deposition rate of each metal under these experimental conditions. To achieve this, we monitor the deposition current during electrodeposition of each metal inside an anodic alumina template of known pore length. The deposition current increases drastically when the pores are completely filled. The deposition rate is determined by calculating the ratio of pore length to pore filling time. According to this calibration, for the spin-valve structure, thicknesses of Ni and Cu layers are estimated to be approximately 40nm each and the Co layer is approximately 20nm thick. Thus, after electrodeposition, we have a two dimensional array of trilayered nanowires vertically standing in an insulating alumina matrix. The template is slightly etched from topside in dilute phosphoric acid in order to expose the tips of the Co layer for electrical contact. The bulk aluminum underneath forms an ohmic contact with electrodeposited Ni and thus serves as a back electrode for the spin valve device (Fig. 3). Contacts pads are made to the top (Co layer) and bottom (Al) using silver paste and gold wires are attached to the contact pads for electrical characterization. The contact areas are ~ 1mm x 1 mm.

Magnetoresistance of this device is measured in a Quantum Design Physical Property Measurement System. Sample temperature is varied in the range 1.9-300K while magnetic field is swept from –70kOe to +70kOe. Applied bias current is 10 $\mu A$.

It must be noted that not all nanowires are electrically contacted from both ends. Reverse polarity etching technique does not completely remove the entire barrier layer from backside as can be seen immediately from Fig. 2 (b). Moreover, not every nanowire tip is exposed from the top. The device resistance is $\sim 300\,\Omega$ which is quite large for an all-metal device. However, this is not a surprise because the conductivities of metal quantum wires can be much less compared to their bulk values[10]. This is because at low dimensions electron mean free path is limited by surface scattering. If we assume that the total conductivity $\sigma$ of a trilayered wire is $\sim 1 \times 10^5 \Omega^{-1} m^{-1}$, which is an order of magnitude smaller than typical thin-film conductivity values of metals[1], then we can estimate the resistance of a single wire using the formula $R = l/(\sigma A)$. With $l$ = 100nm, A = $\pi(50nm)^2/4$, we get $R \sim 509\Omega$. Thus, we can say that very few (~2) wires are electrically connected from both sides, even though we use rather large area contacts. In fact, the wire density is about $10^{10}$ cm$^{-2}$, whereas the contact pads have areas of 1 mm$^2$. Therefore, $10^8$ wires are covered by the contacts but only ~2 of them are actually contacted electrically! This behavior is unique to these structures and was observed before. The current-voltage curve of this device shows linear characteristics (Fig.3(b)) at all temperatures, which indicates that the electrical contacts to the few wires that are connected are ohmic.

## 3. RESULTS AND DISCUSSION

The magnetoresistance traces at three different temperatures (for a fixed bias current of 10 $\mu A$) are shown in Fig. 4(a)-(c). As indicated in Fig. 3(a), the magnetic field is along the length of the nanowires. This coincides with the easy axes of magnetization for the ferromagnetic electrodes. We observe a global positive magnetoresistance which has been reported earlier by similar studies. The background positive magnetoresistance comes about from progressively increasing magnetization of the ferromagnetic contacts with increasing magnetic field strength. In addition, we observe tell-tale magnetoresistance peaks between the coercive fields of the nanomagnets (say |H$_{c,Ni}$| and |H$_{c,Co}$|, with |H$_{c,Ni}$|<|H$_{c,Co}$| in general) which is the characteristic signature of the spin valve effect. In the region |H$_{c,Ni}$|<|H|<|H$_{c,Co}$| magnetizations of

the nanomagnets are *antiparallel* which results in a high device resistance since one contact injects spins of a particular polarization and the other contact blocks these electrons from getting through. This effect manifests itself as the peaks in the magnetoresistance curves. Outside the region |H$_{c,Ni}$|<|H|<|H$_{c,Co}$|, the magnetizations of the two ferromagnetic contacts are parallel resulting in smaller device resistance. This is the basic spin valve effect.

Coercivities of Ni and Co nanomagnets electrodeposited in porous alumina template have been well calibrated in the past[11, 12]. Ref. [12] reports coercivity of electrodeposited cylindrical Co nanomagnets as a function of diameter and length. Co nanodots with 20nm length and 10nm diameter have coercivity ~1000 Oe at room temperature. Also coercivity decreases as dot diameter is increased. Extrapolating this trend, in the present case, where Co nanodots have 50nm diameter (and 20 nm length), coercivity is estimated to be ~700 Oe at room temperature. At lower temperatures we expect to see a somewhat higher value than this estimate. Indeed, in the temperature range 1.9K-100K, we observe |H$_{c,Co}$|~750 Oe. Similarly, for Ni, extrapolating the data in ref. [11] we expect room temperature coercivity of Ni nanodots of 50nm diameter to be ~200 Oe. Thus in the temperature range 1.9K-100K, |H$_{c,Ni}$|~250 Oe is quite likely. Therefore, the magnetizations of the nanomagnets are antiparallel in the range [250 Oe, 750 Oe] in the temperature range 1.9K-100K. We observe magnetoresistance peaks in this field range which confirms that what we observe is indeed the spin valve effect.

It is possible to estimate the spin relaxation length in Cu spacer from the knowledge of the spin valve signal $\Delta R$ which is the change in device resistance from parallel to antiparallel configuration. We follow the model of ref.[1] modified for the classical spin valve geometry:

$$\Delta R = \frac{2\alpha_F^2 \frac{\lambda_N}{\sigma_N A} e^{(-L/2\lambda_N)}}{(M+1)[M\sinh(L/2\lambda_N)+\cosh(L/2\lambda_N)]} \quad (1)$$

where

$M = \frac{\lambda_N \sigma_F (1-\alpha_F^2)}{\lambda_F \sigma_N}$, $\alpha_F = \frac{\sigma_\uparrow - \sigma_\downarrow}{\sigma_\uparrow + \sigma_\downarrow} \equiv$ bulk

current polarization of the ferromagnetic electrodes (assuming they are made of same material), $\sigma_\uparrow(\sigma_\downarrow)$ indicates the spin up (down) conductivity of the ferromagnet, $\sigma_N(\sigma_F)$ denote the total conductivity of the Cu (ferromagnetic) layer, $\lambda_N(\lambda_F)$ is the spin relaxation length in the Cu (ferromagnetic) layer, $L = 40 nm$ is the distance between the two ferromagnetic electrodes and $A$ is the cross sectional area through which current flows into the spin valve device. We make the following reasonable estimates to calculate $\lambda_N$, which is the spin relaxation length in Cu:

(a) $\alpha_F = 0.375$ ("average" spin polarization of Ni (33%) and Co electrodes (42%).
(b) $\sigma_N = 1\times 10^5 \Omega^{-1} m^{-1}$ (electrical conductivity of nanowire Cu)
(c) $\sigma_F = 1\times 10^5 \Omega^{-1} m^{-1}$ ("average" electrical conductivity of the nanowire ferromagnets)
Note that in case of both Cu and the ferromagnets we have assumed a smaller conductivity value (compared to bulk or thin films) in order to incorporate the increased surface roughness scattering which decreases carrier mobility. For simplicity we have assumed same conductivity value for both Cu and the ferromagnets.
(d) $\lambda_F = 5nm$ (spin relaxation length in the ferromagnets, see ref.[13])
(e) $A = \pi(50nm)^2 \times 2/4 = 3.927\times 10^{-15} m^2$ (current carrying cross sectional area of the device, assuming ~ 2 nanowires are electrically connected from both sides; each wire has a diameter of 50 nm)
(f) $L = 40nm$ (thickness of the Cu layer, as discussed earlier)
(g) $\Delta R$ is determined from the magnetoresistance curves (Fig. 4(a)-(c)) for various temperatures.

Using these values we obtain the following estimates of $\lambda_N$ from equation (1). We can also find the spin lifetime $\tau$ using the formula[14] $\lambda_N = \sqrt{\frac{v_F \tau \lambda_m}{3}}$ where $v_F = 1.57 \times 10^6$ m/s is the Fermi velocity in Cu[15] and $\lambda_m \approx 5$ nm[14] is the mean free path (limited by surface roughness scattering) in Cu spacer. Note that this value of $\lambda_m$ is somewhat smaller than the bulk value (~ 50 nm). This is because of increased surface roughness scattering in small dimensions. The results are shown in the following table:

| Temperature (K) | $\Delta R$ ($\Omega$) | $\lambda_N$ (nm) | $\tau$ (fs) |
|---|---|---|---|
| 1.9 | 0.18 | 18 | 123.82 |
| 50 | 0.17 | 16 | 97.83 |
| 100 | 0.17 | 16 | 97.83 |

Comparing these values to the value of spin relaxation length in thin films (ref.[1]), we find that the spin relaxation length has been *reduced* by almost an order of magnitude. This is a consequence of increased surface roughness scattering which decreases the mobility and therefore exacerbates the EY mechanism, while simultaneously quenching the DP mechanism. Thus, what we are measuring is the *spin relaxation rate associated with pure EY mechanism alone.* Typical mean free time in the Cu spacer (determined by surface roughness scattering) is approximately given by $\tau_f = \lambda_m/v_F = 3.1$ fs. Thus, on an average, spin relaxation due to EY mechanism occurs over 30-40 $(=\tau/\tau_f)$ momentum scattering events.

Note from the above table that spin relaxation length is relatively independent of temperature. This is consistent with the EY mechanism. The primary source of the EY relaxation is the surface roughness scattering, which is an elastic scattering event that does not involve phonons. Thus it is expected to be relatively temperature independent.

Fig.5 shows magnetoresistance plot at a higher bias current of 100 $\mu$A. However, the height of the spin valve peak does not change appreciably by this tenfold increase in bias. Surface roughness scattering does not strongly depend on applied bias and hence spin valve signal is expected to be weakly dependent on it at least in 10-100 $\mu$A range. We have not gone past the 100 $\mu$A limit in order to prevent sample damage. We point out that the voltage on the sample at 100 $\mu$A is 100 $\mu$A x 300 $\Omega$ = 30 mV. Therefore, the electric field on the Cu spacer layer is 30 mV/40 nm = 7.5 kV/cm.

Ref.[1] reported a significant decrease in spin relaxation length with an increase in temperature for Cu thin films. This is in sharp contrast with what we find. Given that the DP process has a relatively weak temperature dependence[16] it seems that EY is still the dominant spin relaxation mechanism in thin films. However, in thin films, the primary mobility degradation mechanism is not surface roughness scattering but probably phonon scattering which causes the strong temperature dependence.

In conclusion, we have isolated the EY mechanism in Cu by utilizing a nanowire and studied the spin relaxation rates associated with his mechanism at various temperatures and electric fields.

This work is supported by the US Air Force Office of Scientific Research under grant FA9550-04-1-0261.

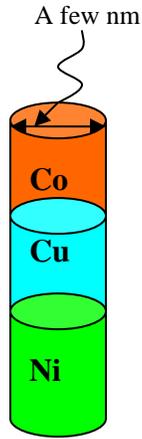

Fig. 1: A nanowire spin valve structure consisting of Co, Cu and Ni.

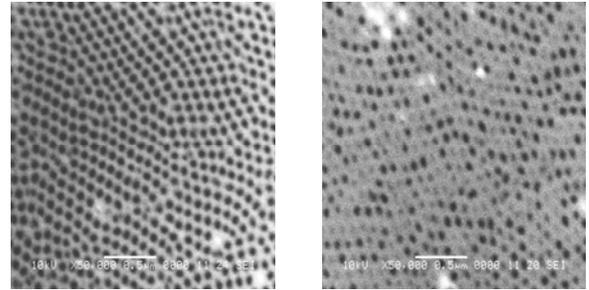

(a)             (b)

Fig. 2: SEM micrographs of alumina template formed by anodization using 3% oxalic acid at 40V dc: (a) top view, showing nominal pore diameter of 50nm (b) bottom view (after removing the bulk alumina) showing through pores obtained as a result of reverse polarity etching.

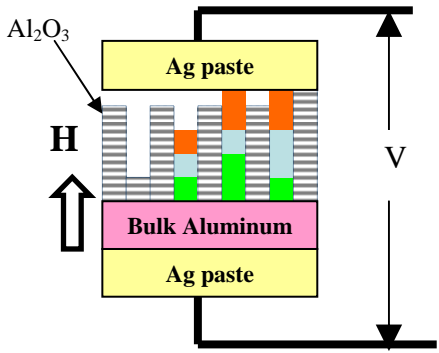

Fig. 3(a): A schematic representation of the all-metal quantum wire spin valve device. Note that not all quantum wires are electrically connected from both sides. Here only two wires are shown connected from both ends.

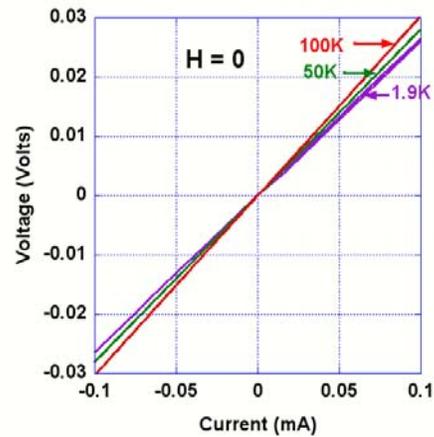

Fig.3(b): The linear current - voltage characteristics of the device shown in Fig. 3 (a) indicate that the heterojunctions as well as the electrical contacts are ohmic in nature.

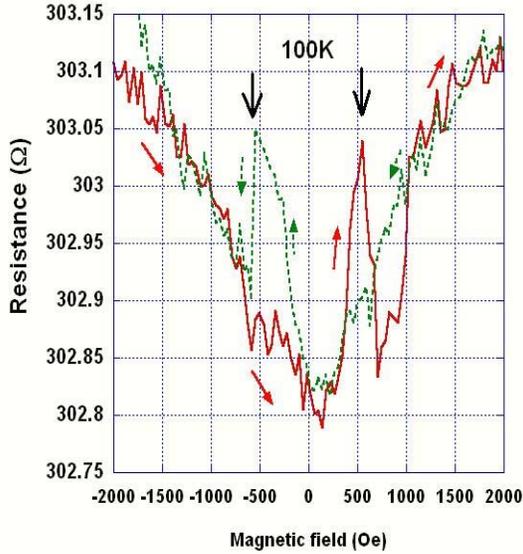

Fig. 4(a): Magnetoresistance (MR) characteristics at 100K. Bias current is 10 $\mu A$. Vertical arrows show the spin valve peaks.

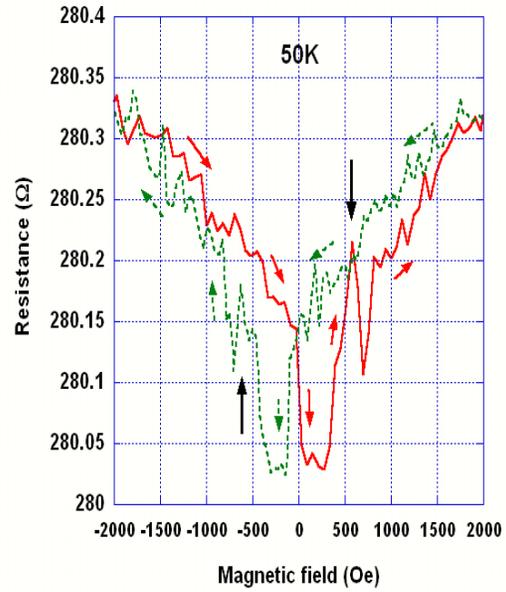

Fig.4 (b): MR characteristics at 50K and 10 $\mu A$ bias current. Vertical arrows show the spin valve peaks.

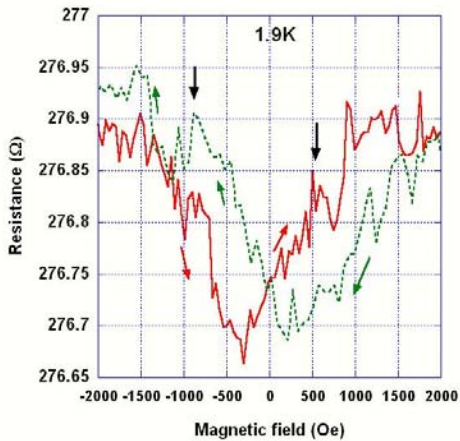

Fig. 4(c): MR characteristics at 1.9K and 10 $\mu A$ bias current. Vertical arrows show the spin valve peaks.

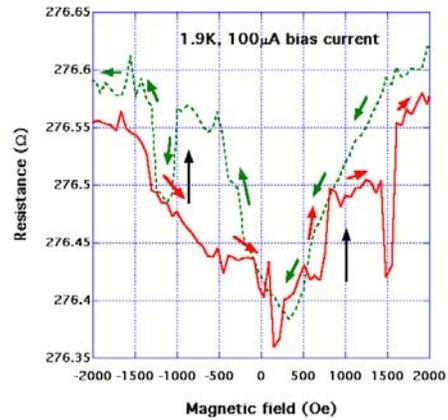

Fig.5: MR characteristics at 1.9K and 100 $\mu A$ bias current. Vertical arrows show the spin valve peaks.